# Wetting of an annular liquid in a capillary tube

Cunjing Lv[1*] and Steffen Hardt [2]

[1] Department of Engineering Mechanics, Tsinghua University, 100084 Beijing, China

[2] Institute for Nano- and Microfluidics, Technische Universität Darmstadt, 64287 Darmstadt, Germany

## ABSTRACT

In this paper, we systematically investigate the wetting behavior of a liquid ring in a cylindrical capillary tube. We obtain analytical solutions of the axisymmetric Young-Laplace equation for arbitrary contact angles. We find that, for specific values of the contact angle and the volume of the liquid ring, two solutions of the Young-Laplace equation exist, but only the one with the lower value of the total interfacial energy corresponds to a stable configuration. The transition to an unstable configuration is characterized by specific critical parameters such as the liquid volume, throat diameter etc. Beyond the stable regime, the liquid ring transforms into a plug. Based on numerical simulations, we also discuss the transition of an axisymmetric ring into non-axisymmetric configurations. Such a transition can be induced by the Rayleigh-Plateau instability of a slender liquid ring. The results are presented in terms of a map showing the different liquid morphologies as a function of the parameters governing the problem.

[*] To whom correspondence should be address. Email: cunjinglv@tsinghua.edu.cn

## INTRODUCTION

Wetting of solids by liquids is ubiquitous in nature and industry. Depending on the boundary conditions, the volumetric forces and the geometry of the surface, various liquid morphologies are observed (1-3), for example spherical-cap drops on a solid, liquid bridges confined between solids (3, 4), barrel-shaped and clamshell drops on fibers (5), liquid rings on a substrate (6), as well as liquid puddles with a hole under the centrifugal force (7), etc. Moreover, in these cases, the liquid could be either in a stable static equilibrium configuration or potentially lose stability with morphological transitions. Very recently, Bostwick *et al.* (3) reviewed the hydrodynamic stability of capillary surface subject to various constraints, such as volume conservation, contact-line boundary conditions and the geometry of the supporting surface. Owing to its omnipresence and practical relevance in such fields as capillary condensation (8, 9), fluid transport (10, 13, 14), and thermocapillarity (11, 12), wetting under confinement has received enormous attention. Especially liquid surfaces that take the shape of a (deformed) cylinder have been in the focus of intense research efforts. In this paper, we attempt to construct a rigorous framework to characterize the wetting of an annular liquid in a capillary tube.

When a system becomes sufficiently small and the effects of gravity can be ignored, cylindrical liquid surfaces with constant surface tension tend to minimize their free energy by adopting unduloidal shapes. This means that the mean curvature of a surface is constant. In this context, an exact analytical solution for the interface profile was found (15-18). There is extensive literature on the stability of cylindrical liquid surfaces (3, 19-27). An infinitely extended liquid cylinder is always unstable under perturbations of sufficiently long wavelength (28). Everett (19) obtained unduloidal surfaces for a liquid completely wetting the interior of a cylindrical tube. They also obtained the maximum volume corresponding to the closure of the tube by the liquid. Gauglitz and Radke (21) investigated the dynamics of a liquid film inside a cylindrical capillary and its transition into collars or plugs. Their numerical calculations suggest that a liquid collar becomes unstable and evolves into a plug when its radius reaches a critical value. By employing asymptotic methods, Jensen (24) quantified the gravitational effects on the quasi-steady evolution of a liquid wetting the interior of a vertical cylindrical tube. He analyzed the evolution of structures such as collars and plugs. Based on the finite-element method, Collicott (25) studied the possible morphologies of a liquid in a circular tube without gravity at all contact angles ranging from 0° to 180°. They especially focused on non-axisymmetric configurations. Slobozhanin (22) examined the axisymmetric equilibrium configurations and stability of a liquid that partially fills a cylindrical container with planar ends, and analyzed the roles of the contact angle, liquid volume and Bond number. In a very recent work, Chen (27) systematically investigated the instabilities of axisymmetric and non-axisymmetric fluid threads confined in a channel and separated from the walls by a lubrication layer. They put their focus on the effects of confinement on the stability of the threads.

Although a lot of work has already been done, there is still a need for rigorous theories describing the liquid morphologies and the transitions between them over a broad parameter range, for example with respect to the contact angle on the solid surface. One challenge is to find a sufficiently general solution of the Young-Laplace equation for the stable static equilibrium configuration and to determine the critical parameters of the instability. In that context, up to now analytical solutions were obtained for some complete wetting states (19, 20), or numerical methods (22, 24, 25, 27) were employed, or the conclusions were built on approximations (21) in which the radial coordinate of the liquid surface is supposed to be some function of the axial coordinate. The objective of the present paper is the solution of the Young-Laplace equation for an axisymmetric liquid volume at zero gravity in a cylindrical capillary with an arbitrary contact angle. Based on that, we obtain the criteria characterizing the transition of a liquid ring to a liquid plug. The studies are complemented by numerical simulations that especially aim at situations where the axial symmetry is broken.

## STATIC PROFILE OF A LIQUID RING IN A CAPILLARY TUBE

### General solution

Considering capillary tubes with a diameter significantly smaller than the capillary length $l_c = (\sigma/\rho g)^{1/2}$ (~ 2.73 mm for water under standard conditions), we ignore the influence of gravity, so the pressure $P_L$ inside the liquid is constant. We denote $\sigma$, $\rho$ and $g$ the liquid-vapor surface tension, the liquid mass density and the gravitational acceleration, respectively. Taking the atmospheric pressure $P_0$ as constant, we consider the Young-Laplace equation

$$2H = \Delta P = P_L - P_0 \qquad (1)$$

to compute the shape of the liquid surface, where the mean curvature is given by

$$2H = \frac{1}{R_1} + \frac{1}{R_2} = \text{Constant}. \qquad (2)$$

$R_1$ and $R_2$ are the two principle radii of curvature at a point on the liquid surface (see Figure 1a). We assume that the inner wall of the capillary is perfectly smooth, and a constant, well-defined contact angle $\theta$. Especially, this means that any contact-angle hysteresis is neglected.

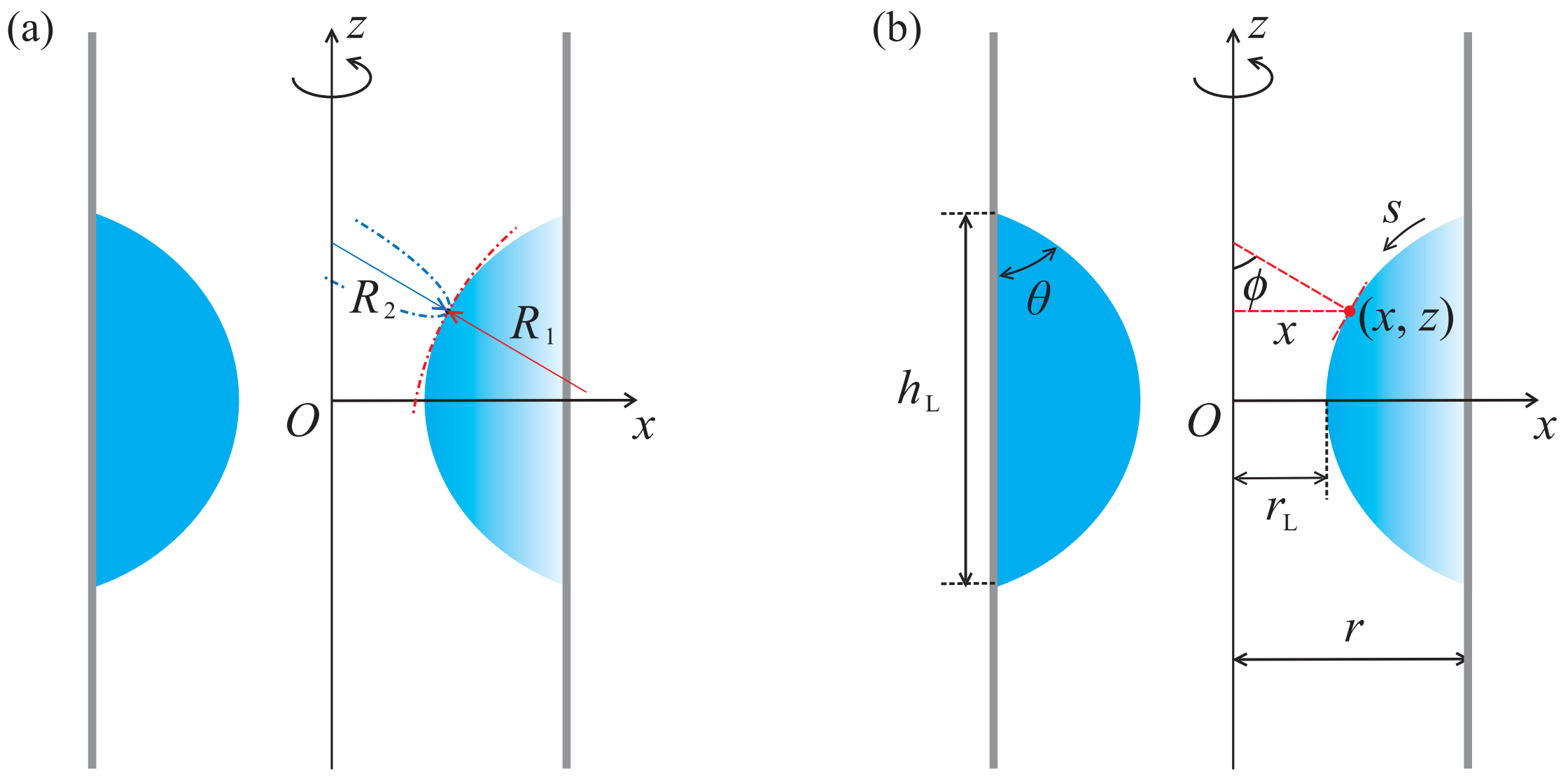

**Figure 1.** Schematic of a liquid ring in a capillary tube. For the mathematical model, a cylindrical coordinate system is chosen. (a) The two principle radii of curvature $R_1$ (red) and $R_2$ (blue) at a specific point of the liquid surface are indicated by circles. (b) Definition of geometric parameters. $x$ and $z$ are the radial and axial coordinates, respectively. $r$ and $r_L$ denote the inner radius of the capillary tube and the radius of the liquid throat, respectively. $h_L$ denotes the height of the solid-liquid contact area. $\theta$ is the contact angle of the liquid. $s$ represents the arc length parameter of the surface profile. $\phi$ is the angle between the surface normal at an arbitrary point $(x, z)$ and the axis of revolution (i.e. $z$-axis).

On the basis of the geometric parameters defined in Figure 1 and elementary differential geometry (29), for a certain point on the liquid surface, we obtain: $(1/R_1) = \mathrm{d}\phi/\mathrm{d}s = \cos\phi\,(\mathrm{d}\phi/\mathrm{d}x)$, and $(1/R_2) = \sin\phi/x$. As a result, $2H$ can be expressed as

$$2H = \cos\phi\frac{\mathrm{d}\phi}{\mathrm{d}x} + \frac{\sin\phi}{x} = \frac{1}{x}\frac{\mathrm{d}}{\mathrm{d}x}\left(x\sin\phi\right), \qquad (3)$$

A first integral of Eq. (3) leads to

$$x\sin\phi = Hx^2 + c_0, \qquad (4)$$

in which $c_0$ is an integration constant. To solve Eq. (4) and determine the two unknown constants $H$ and $c_0$, we have to consider the following two boundary

conditions

$$\text{At } x_1 = r,\ z_1 = \frac{h_{\mathrm{L}}}{2}: \quad \phi_1 = \frac{\pi}{2} - \theta, \tag{5}$$

$$\text{At } x_2 = r_{\mathrm{L}},\ z_2 = 0: \quad \phi_2 = \frac{\pi}{2}. \tag{6}$$

Substituting Eqs. (5) and (6) into Eq. (4), we find

$$2H = 2\left(\frac{r\cos\theta - r_{\mathrm{L}}}{r^2 - r_{\mathrm{L}}^2}\right), \tag{7}$$

$$c_0 = \left(\frac{r - r_{\mathrm{L}}\cos\theta}{r^2 - r_{\mathrm{L}}^2}\right)\cdot r r_{\mathrm{L}}. \tag{8}$$

When inserting Eqs. (7) and (8) into Eq. (4), we obtain

$$x\sin\phi = \left(\frac{r\cos\theta - r_{\mathrm{L}}}{r^2 - r_{\mathrm{L}}^2}\right)x^2 + \left(\frac{r - r_{\mathrm{L}}\cos\theta}{r^2 - r_{\mathrm{L}}^2}\right)\cdot r r_{\mathrm{L}}. \tag{9}$$

For a given system, $r$ and $\theta$ are known parameters, and when a static liquid ring is formed, $r_{\mathrm{L}}$ has a specific value, so the surface profile can be determined by Eq. (9) in terms of $x$ and $\phi$. Up to this point, the mathematical formalism closely follows that of Carroll (30).

Next, by using $\mathrm{d}z/\mathrm{d}x = \tan\phi$ and $\tan\phi = \sin\phi/(1 - \sin^2\phi)^{1/2}$, we can eliminate $\phi$ in Eq. (9). After a rather tedious calculation we obtain

$$\frac{\mathrm{d}z}{\mathrm{d}x} = \frac{x^2\left(r_{\mathrm{L}} - r\cos\theta\right) + r r_{\mathrm{L}}\left(r_{\mathrm{L}}\cos\theta - r\right)}{\left[x^2\left(r_{\mathrm{L}} - r\cos\theta\right)^2 - r^2\left(r_{\mathrm{L}}\cos\theta - r\right)^2\right]^{1/2}\left(r_{\mathrm{L}}^2 - x^2\right)^{1/2}}. \tag{10}$$

By substituting $a = (r\cos\theta - r_L)/(r - r_L\cos\theta)$, we rewrite Eq. (10) in a simpler form:

$$\frac{dz}{dx} = \frac{ax^2 + r \cdot r_L}{\left[\left(x^2 - r_L^2\right)\left(r^2 - a^2x^2\right)\right]^{1/2}}. \tag{11}$$

Note that in Carroll's paper (30), $dz/dx = -\tan\phi$ and $a = (r_L\cos\theta - r)/(r_L - r\cos\theta)$, which is very different from our work because of the completely different liquid morphologies in these two studies, which in turn makes the following calculations substantially different from those reported in Carroll's work (30). By introducing a new variable $\varphi$ and defining $k = [1 - (a\,r_L/r)^2]^{1/2}$, we obtain the following identity

$$\left(ax\right)^2 = r^2\left(1 - k^2\sin^2\varphi\right). \tag{12}$$

That way we can transform $z = z(x)$ into $z = z(\varphi)$. Based on that, the boundary conditions Eqs. (5) and (6) are rewritten into

$$\text{At } x_1 = r,\ z_1 = \frac{h_L}{2}:\quad \varphi_1 = \arcsin\sqrt{\frac{1-a^2}{k^2}}, \tag{13}$$

$$\text{At } x_2 = r_L,\ z_2 = 0:\quad \varphi_2 = \frac{\pi}{2}. \tag{14}$$

In what follows, we calculate various quantities of interest such as $h_L$, the solid-liquid and liquid-vapor interfacial areas $A_{SL}$ and $A_{LV}$, the interfacial energy $U$, and the liquid volume $V_L$. The interfacial energy $U$ is defined by $U = \sigma A_{LV} + (\sigma_{SL} - \sigma_{SV})A_{SL} = \sigma(A_{LV} - A_{SL}\cos\theta)$, denoting $\sigma_{SL}$ and $\sigma_{SV}$ the solid-liquid and solid-vapor interfacial tensions, respectively. First, combining Eqs. (11) and (12) leads to

$$
\begin{aligned}
h_{\mathrm{L}} &= 2\int_{x_1}^{x_2}\left(\frac{\mathrm{d}z}{\mathrm{d}x}\right)\mathrm{d}x = 2\int_{\varphi_1}^{\varphi_2}\left(\frac{\mathrm{d}z}{\mathrm{d}x}\cdot\frac{\mathrm{d}x}{\mathrm{d}\varphi}\right)\mathrm{d}\varphi \\
&= 2\int_{\varphi_1}^{\varphi_2}\left(\frac{r}{a}\sqrt{1-k^2\sin^2\varphi}+\frac{r_{\mathrm{L}}}{\sqrt{1-k^2\sin^2\varphi}}\right)\mathrm{d}\varphi\,, \\
&= 2\left[\frac{r}{a}E(\varphi,k)+r_{\mathrm{L}}F(\varphi,k)\right]\Bigg|_{\varphi_1}^{\varphi_2=\frac{\pi}{2}}
\end{aligned}
\tag{15}
$$

in which d$z$/d$x$ and d$x$/d$\varphi$ are computed using Eqs. (11) and (12), respectively. Eq. (15) leads to

$$
A_{\mathrm{SL}} = 2\pi r h_{\mathrm{L}} = 4\pi r\left[\frac{r}{a}E(\varphi,k)+r_{\mathrm{L}}F(\varphi,k)\right]\Bigg|_{\varphi_1}^{\varphi_2=\frac{\pi}{2}}. \tag{16}
$$

Similarly, we obtain

$$
\begin{aligned}
A_{\mathrm{LV}} &= 2\int_{s_1}^{s_2}(2\pi x)\,\mathrm{d}s = 4\pi\int_{\varphi_1}^{\varphi_2}\left(x\frac{\mathrm{d}s}{\mathrm{d}\varphi}\right)\mathrm{d}\varphi \\
&= 4\pi\left(\frac{r}{a}\right)\left(\frac{r}{a}+r_{\mathrm{L}}\right)E(\varphi,k)\Big|_{\varphi_1}^{\varphi_2=\frac{\pi}{2}}
\end{aligned}\,, \tag{17}
$$

$$
\begin{aligned}
U &= \sigma\left(A_{\mathrm{LV}} - A_{\mathrm{SL}}\cos\theta\right) \\
&= 4\pi\sigma\left[\left(\frac{r}{a}\right)\left(\frac{r}{a}+r_{\mathrm{L}}-r\cos\theta\right)E(\varphi,k)\Big|_{\varphi_1}^{\varphi_2=\frac{\pi}{2}} - r\cdot r_{\mathrm{L}}\cos\theta\, F(\varphi,k)\Big|_{\varphi_1}^{\varphi_2=\frac{\pi}{2}}\right],
\end{aligned} \tag{18}
$$

$$
\begin{aligned}
V_{\mathrm{L}} &= \pi r^2 h_{\mathrm{L}} - 2\int_{z_1}^{z_2} \pi x^2 \mathrm{d}z = \pi r^2 h_{\mathrm{L}} - 2\pi \int_{z_1}^{z_2} \left( x^2 \frac{\mathrm{d}z}{\mathrm{d}\varphi} \right) \mathrm{d}\varphi \\
&= 2\pi r^2 \left[ \frac{r}{a} E(\varphi,k) + r_{\mathrm{L}} F(\varphi,k) \right] \Bigg|_{\varphi_1}^{\varphi_2 = \frac{\pi}{2}} \\
&- \frac{2\pi}{3} \left( \frac{r_{\mathrm{L}}^3}{a} \right) \left( \frac{r}{r_{\mathrm{L}}} \right) \left\{ \left[ \frac{2}{a^2} \left( \frac{r}{r_{\mathrm{L}}} \right)^2 + \frac{3}{a} \left( \frac{r}{r_{\mathrm{L}}} \right) + 2 \right] E(\varphi,k) \Big|_{\varphi_1}^{\varphi_2 = \frac{\pi}{2}} \right. \\
&\left. - F(\varphi,k) \Big|_{\varphi_1}^{\varphi_2 = \frac{\pi}{2}} - \frac{r}{r_{\mathrm{L}}} \sqrt{1 - a^2} \sqrt{\left( \frac{r}{r_{\mathrm{L}}} \right)^2 - 1} \right\}
\end{aligned}, \tag{19}
$$

in which $\mathrm{d}s = [1 + (\mathrm{d}z/\mathrm{d}x)^2]^{1/2}\, \mathrm{d}x$ and $\mathrm{d}z/\mathrm{d}\varphi = (\mathrm{d}z/\mathrm{d}x)(\mathrm{d}x/\mathrm{d}\varphi)$ are employed. $F(\varphi, k)$ and $E(\varphi, k)$ are the incomplete elliptic integrals of the first and the second kind (31), respectively. Therefore, by employing the transformation of Eq. (12), we obtain the expressions of $h_{\mathrm{L}}$, $A_{\mathrm{SL}}$, $A_{\mathrm{LV}}$, $U$ and $V_{\mathrm{L}}$ as functions of $r$, $\theta$ and $r_{\mathrm{L}}$, taking into account that $\varphi_1$ is a function of $r$, $\theta$ and $r_{\mathrm{L}}$.

We also obtain the shape of the liquid surface. For any point $z_i(x_i) = z_i(\varphi_i)$ at the upper half of the surface we have

$$
\begin{aligned}
z_i(x_i) &= \int_{x_1}^{x_i} \left( \frac{\mathrm{d}z}{\mathrm{d}x} \right) \mathrm{d}x = \int_{\varphi_1}^{\varphi_i} \left( \frac{\mathrm{d}z}{\mathrm{d}x} \cdot \frac{\mathrm{d}x}{\mathrm{d}\varphi} \right) \mathrm{d}\varphi \\
&= \int_{\varphi_1}^{\varphi_i} \left( \frac{r}{a} \sqrt{1 - k^2 \sin^2 \varphi} + \frac{r_{\mathrm{L}}}{\sqrt{1 - k^2 \sin^2 \varphi}} \right) \mathrm{d}\varphi, \\
&= \left[ \frac{r}{a} E(\varphi,k) + r_{\mathrm{L}} F(\varphi,k) \right] \Bigg|_{\varphi_1}^{\varphi_i}
\end{aligned} \tag{20}
$$

in which $\varphi_i \in [\varphi_1, \varphi_2]$, corresponding to $x_i \in [x_1, x_2]$ and $z_i \in [z_1, z_2]$. Eq. (20) suggests that we can also obtain the lower half of the shape of the liquid surface because of symmetry.

In a practice, the inner radius $r$ of the capillary tube and the contact angle $\theta$ are known,

so for a given value of $r_L$, the other quantities such as $2H$, $h_L$, $A_{SL}$, $A_{LV}$, $U$ and $V_L$ are determined through Eqs. (7), (15)-(19). Otherwise, if we know $h_L$, on the basis of Eq. (15) in which $h_L = h_L(r, \theta, r_L)$, we can determine $r_L$ and then the other quantities through Eqs. (7), (16)-(19). Generally speaking, considering the functions $H = H(r, \theta, r_L)$, $h_L = h_L(r, \theta, r_L)$, $A_{SL} = A_{SL}(r, \theta, r_L)$, $A_{LV} = A_{LV}(r, \theta, r_L)$, $U = U(r, \theta, r_L)$ and $V_L = V_L(r, \theta, r_L)$, if one of these is known, we can determine $r_L$ and then determine the other quantities. The profile of the liquid ring can be found as well. Based on the above equations, knowing $r_L$ is the simplest option to determine all other quantities. For the other cases, numerical methods may be employed (for example, to determine $r_L$) because there are no explicit expressions of the elliptic integrals. In fact, the surface profile belongs to the family of unduloidal curves (15), which describe a number of capillary surfaces (16-18). However, to our knowledge, up to now general closed-form expressions for the quantities of interest such as $2H$, $h_L$, $A_{SL}$, $A_{LV}$, $U$, $V_L$ have not been reported, particularly for arbitrary contact angles $\theta$.

**Special case $\theta = 0°$**

In the following we would like to shine more light on the special case $\theta = 0°$, because of its importance for realistic wetting phenomena in capillary tubes. Examples are the transport of confined foam lamellae (13), the influence of capillary and Marangoni forces on transport phenomena in microgravity (11, 12), the pinch-off of bubbles (32), or the instability of confined fluid threads (27). In many of these cases, a completely wetting liquid, i.e. $\theta = 0°$ can be assumed.

When $\theta = 0°$, the problem simplifies considerably. Based on the above definitions, we obtain $a = 1$ and $k = [1 - (r_L/r)^2]^{1/2}$. As a result, the boundary conditions transform into (i) At $x = r$, $z = h_L/2$: $\phi_1 = \pi/2$, $\varphi_1 = 0$; and (ii) At $x = r_L$, $z = 0$: $\phi_2 = \pi/2$, $\varphi_2 = \pi/2$. Moreover, the equations for the quantities of interest transform into

$$2H = \frac{2}{r + r_{\mathrm{L}}}, \tag{21}$$

$$h_{\mathrm{L}} = 2\left[r \cdot E(k) + r_{\mathrm{L}} \cdot K(k)\right], \tag{22}$$

$$A_{\mathrm{SL}} = 4\pi r\left[r \cdot E(k) + r_{\mathrm{L}} \cdot K(k)\right], \tag{23}$$

$$A_{\mathrm{LV}} = 4\pi r\left(r + r_{\mathrm{L}}\right)E(k), \tag{24}$$

$$U = 4\pi\sigma\left\{r \cdot r_{\mathrm{L}}\left[E(k) - K(k)\right]\right\}, \tag{25}$$

$$\begin{aligned} V_{\mathrm{L}} &= 2\pi r^2\left[r \cdot E(k) + r_{\mathrm{L}} K(k)\right] \\ &\quad - \frac{2\pi}{3} r_{\mathrm{L}}^3 \left(\frac{r}{r_{\mathrm{L}}}\right)\left\{\left[2\left(\frac{r}{r_{\mathrm{L}}}\right)^2 + 3\left(\frac{r}{r_{\mathrm{L}}}\right) + 2\right]E(k) - K(k)\right\}, \end{aligned} \tag{26}$$

in which $K(k)$ and $E(k)$ are the complete elliptic integrals of the first and the second kind (31), respectively.

In Figure 2, we display two classes of surfaces profiles obtained using the above results. The profiles are represented in non-dimensional coordinates, using the inner diameter $d$ ($d = 2r$) of the capillary tube as the length scale. In Figure 2a, the contact angle is fixed at $\theta = 0°$, and the curves correspond to different diameters of the liquid throat $d_{\mathrm{L}} = 2r_{\mathrm{L}}$ ranging from $d_{\mathrm{L}}/d = 0.8$ to $d_{\mathrm{L}}/d = 0.05$. For comparison, Figure 2b shows surface profiles for different contact angles ranging from $\theta = 0°$ to $120°$, when the volume is fixed at $V_{\mathrm{L}}/r^3 = 1.25$.

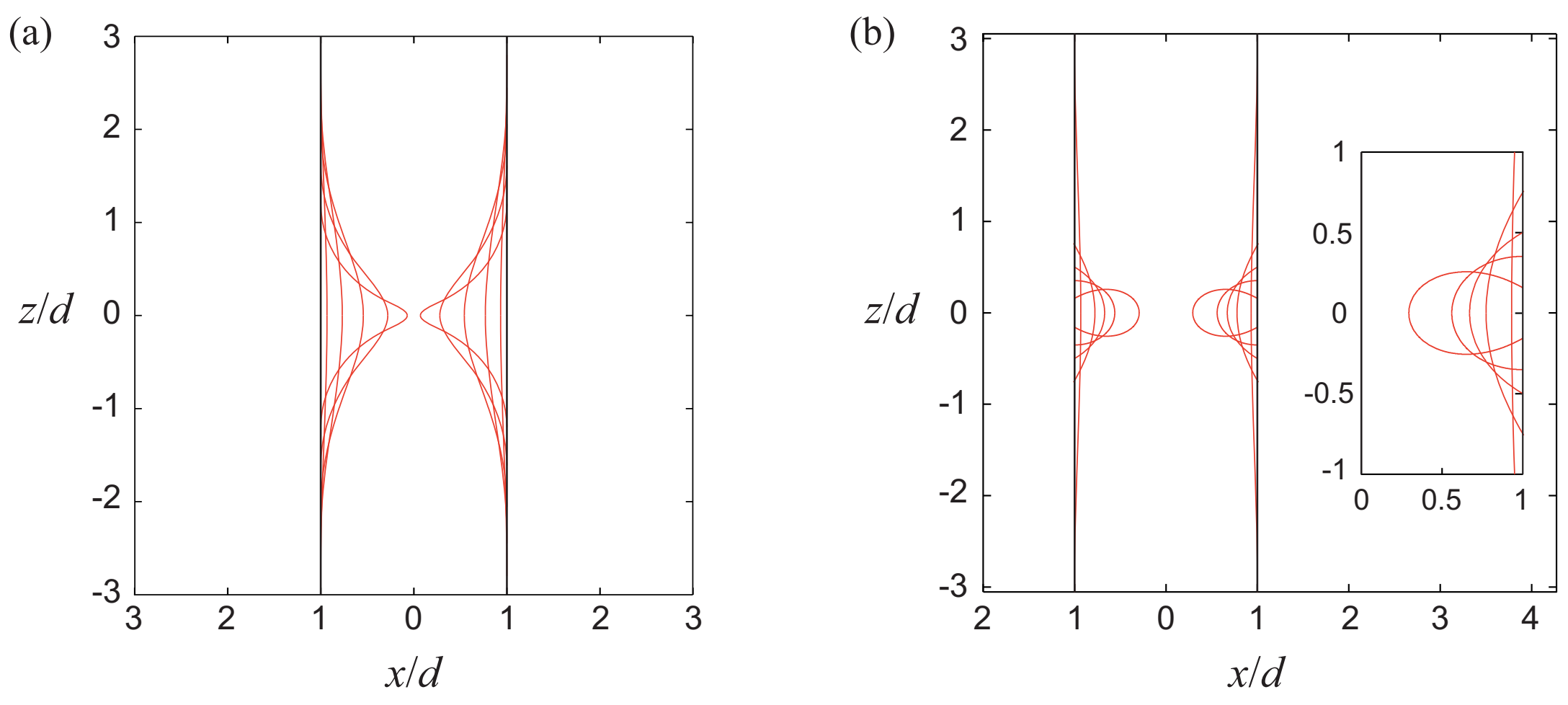

**Figure 2.** Axisymmetric liquid surface profiles in capillary tubes. The two black vertical lines represent the inner wall of the capillary tube. (a) For a fixed contact angle $\theta = 0°$ and for different diameters of the liquid throat, i.e. $d_L/d$ = 0.8, 0.6, 0.4, 0.2, 0.05, respectively. (b) For a fixed liquid volume $V_L/(d/2)^3 = 1.25$ and different contact angles, i.e. $\theta$ = 0°, 30°, 60°, 90°, 120°, respectively. The inset is a blow-up to make details better visible.

As apparent from Figure 2a, solutions of the Young-Laplace equation exist over the full range of liquid throat diameters, i.e. $d_L/d \in [0, 1]$. However, the sheer existence of a solution does not allow drawing conclusions about its stability, an aspect that we will address in the next section.

## MORPHOLOGY TRANSITIONS

### Criteria for the transition

Figure 3, on the basis of the analytical solutions we have obtained, displays the dependencies of the liquid volume $V_L$ and the interfacial energy $U$ on the liquid throat diameter $d_L$ for various contact angles $\theta$. Moreover, in order to check the theoretical results, we performed finite-element method (FEM) simulations (the dots) using the public domain software package *Surface Evolver* (SE) (33), which was developed to compute static capillary surfaces. For convenience, the results are nondimensionalized using the scales $r$ and $\sigma$. The curves indicate that the theoretical results agree very well with the results of the SE calculations.

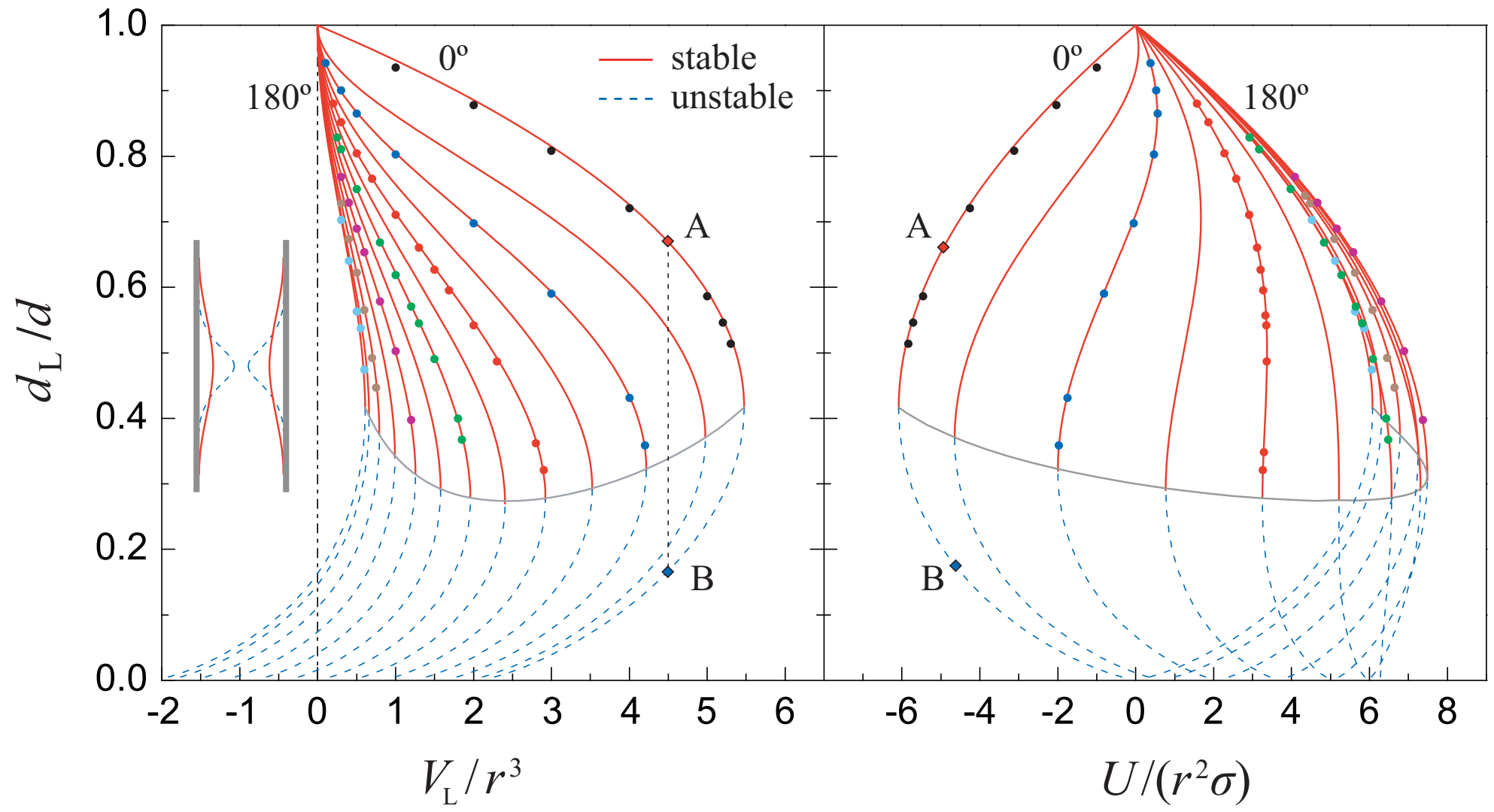

**Figure 3.** Dependence of $V_L$ and $U$ on $d_L$ expressed in dimensionless form. The contact angle ranges from 0° to 180°, with increments of 15° for the theoretical (curves) and of 30° for simulation (dots) results. Dots with the same color in the left and the right panels represent the same value of $\theta$. The solid red and dashed blue dotted curves represent the stable and unstable regimes, respectively, with the solid grey curves representing the boundaries between the stable and unstable regime. The dashed-dotted line represents $V_L = 0$, and the points A and B connected by the dotted line represent two wetting states (see the inset in the left panel) with the same value of $V_L$ but different values of $U$ (see the right panel).

The following features of the solution become apparent. First, Figure 3 represents a multiple solution map of the surface profile. In order to illustrate that, we take $\theta = 0°$ as an example. A given liquid volume $V_L/r^3 \approx 4.5$ corresponds to two different surface profiles which are plotted in the inset of the left panel. These two profiles correspond to $d_{LA}/d \approx 0.67$ (point A) and $d_{LB}/d \approx 0.17$ (point B), respectively. However, the right panel suggests that the dimensionless surface energy corresponding to A ($U_A/(r^2\sigma) \approx -4.85$) is lower than that corresponding to B ($U_B/(r^2\sigma) \approx -4.52$). Generally, for a specific value of $\theta$, a given volume $V_L$ corresponds to two values of $d_L$, i.e., $V_L(d_{LA}) = V_L(d_{LB})$, but these two volumes correspond to two different values of the surface energy $U(d_{LA}) < U(d_{LB})$, with $d_{LA} > d_{LB}$. Based on a stability analysis (18) (to be further detailed in the next subsection), we know that the configuration with the lower value of $U$ is stable. This means that under a small perturbation, configuration B would transform into A, which is the only one found in experiments (34). We distinguish the stable and unstable regions by using solid red and dotted blue lines, respectively. Second, the left panel of Figure 3 also indicates that a liquid ring (for arbitrary $\theta$) has a maximum volume $V_{Lc}$, corresponding to a critical throat diameter defined as $d_{Lc}$. This means that when the volume of the liquid ring starts to grow from $V_L = 0$, a static equilibrium ring cannot exist for $V_L > V_{Lc}$. In other words, a liquid ring with $V_L \leq V_{Lc}$ is a stable wetting state, but when $V_L > V_{Lc}$, the liquid assumes a different morphology. $d_{Lc}$ vs. $V_{Lc}$ is represented by the grey curve in the left panel of Figure 3. Moreover, the grey curve in the right panel represents $U = U(d_{Lc})$. In the next section, we will verify that, for a specific value of $\theta$, the stability criterion corresponding to the left panel is the same as the one corresponding to the right panel, and we will rigorously determine the value of the critical parameter $d_{Lc}$. For completeness, the negative volumes appearing in the left panel of Figure 3 represent actual solutions of the Young-Laplace equation that do not bear any physical significance.

## Critical parameters of the transition

For a given capillary tube (with specific values of $r$ and $\theta$), we are particularly interested in the value of $d_{Lc}$. At this critical point, a morphology transition will happen that is of relevance for many applications where small liquid volumes are confined in a capillary (24, 26, 29, 32). Langbein (18) proved a theorem called "Minimum-volume condition", which is very useful for investigating the stability limits of capillary surfaces. Within a family of solutions of the Young-Laplace equation, an instability is reached when the liquid volume exhibits a minimum, maximum or saddle point as a function of the capillary pressure. Thus, if the volume is plotted versus the pressure or any other independent parameter (see Figure 3 and Appendix A), the stability limits show up (18). Based on this rationale, the instability occurs when

$$\frac{\mathrm{d}V_{\mathrm{L}}}{\mathrm{d}\Delta P}=0, \tag{27}$$

or equivalently when

$$\frac{\mathrm{d}V_{\mathrm{L}}}{\mathrm{d}r_{\mathrm{L}}}=0. \tag{28}$$

Based on Eqs. (7) and (19), the solutions of Eqs. (27) and (28) can be obtained. For convenience, we report the detailed calculations in Appendix A. Through either Eq. (27) or Eq. (28), we can determine $d_{\mathrm{Lc}}$, and subsequently other critical quantities such as $2H_{\mathrm{c}}$, $h_{\mathrm{Lc}}$, $A_{\mathrm{SLc}}$, $A_{\mathrm{LVc}}$, $U_{\mathrm{c}}$ and $V_{\mathrm{Lc}}$. Moreover, when we evaluate the criterion $\mathrm{d}U/\mathrm{d}r_{\mathrm{L}} = 0$ (see Appendix A), we obtain the same values of the critical parameters as with Eqs. (27) and (28) and the same stability limits as in Figure 3. However, we have to emphasize that for some $\theta$, in addition to the critical point ($d_{\mathrm{Lc}}$, $U_{\mathrm{c}}$) the criterion $\partial U/\partial r_{\mathrm{L}} = 0$ gives a second solution $d_{\mathrm{L}}$. Discussing the relevance of the second solution, however, is beyond the scope of this paper.

Again, when $\theta = 0°$, computing the critical parameters becomes significantly simpler than in the general case. Then the equation to be solved reads $2E(k) - K(k) = 0$ (see the Supplemental Material). As a consequence, we obtain $k = 0.909$, and the dimensionless critical parameters are $2H_{\mathrm{c}}r = 1.411$, $d_{\mathrm{Lc}}/d = 0.417$, $U_{\mathrm{c}}/(r^2\sigma) = -6.081$, $V_{\mathrm{c}}/r^3 = 5.471$, and $h_{\mathrm{Lc}}/r = 4.257$.

**Morphology transitions in a broader context**

Up to now, the analysis was focused on axisymmetric liquid volumes topologically equivalent to a ring. We have identified a morphology transition, but have not yet discussed the liquid configurations beyond the stability limit. Also, we have taken axisymmetry for granted, but have not considered a potential transition into non-axisymmetric configurations. In this subsection we extend the scope of our analysis to different liquid morphologies.

The criteria for the morphology transition formulated using Eq. (27) or Eq. (28) describe the transition from a liquid ring to a liquid plug, i.e. an axisymmetric liquid volume extending over the entire cross section of the capillary. In reality, one could imagine that there are other types of non-axisymmetric morphologies. If the ring is slender enough (with a suitable value of $\theta$), azimuthal instabilities would occur, and the ring would collapse into satellite drops, arising from the Rayleigh-Plateau instability (35). To study the corresponding liquid morphologies, *Surface Evolver* simulations were performed. Figure 4 is complementary to the left panel of Figure 3 and contains the regions in parameter space where, according to *Surface Evolver*, non-axisymmetric morphologies occur. The *Surface Evolver* data in those regions where axisymmetric configurations are found are indicated by dots. In addition to that, we indeed find a region (green color) where the Rayleigh-Plateau instability lets a slender liquid ring decay into droplets, which is delineated from the stable axisymmetric configurations by the black curve. We use schematic drawings to indicate the wetting states in the two different regions (note that the schematic in the green region just represents the instability, but not the exact instability mode). Therefore, in the green region an axisymmetric liquid ring cannot exist as a stable equilibrium configuration. Making a prediction about the final morphologies obtained in the green region is quite challenging. As a consequence of the inevitable contact angle hysteresis of real surfaces, local imperfections of the capillary wall will, in combination with the dispersion relation of the Rayleigh-Plateau instability, determine the distribution of droplets remaining at the surface. We expect that in the case of significant contact-angle hysteresis, the droplet distribution will be much less uniform than in the case of liquid cylinder (without three-phase contact line). An analysis of this complex scenario is beyond the scope of the present paper.

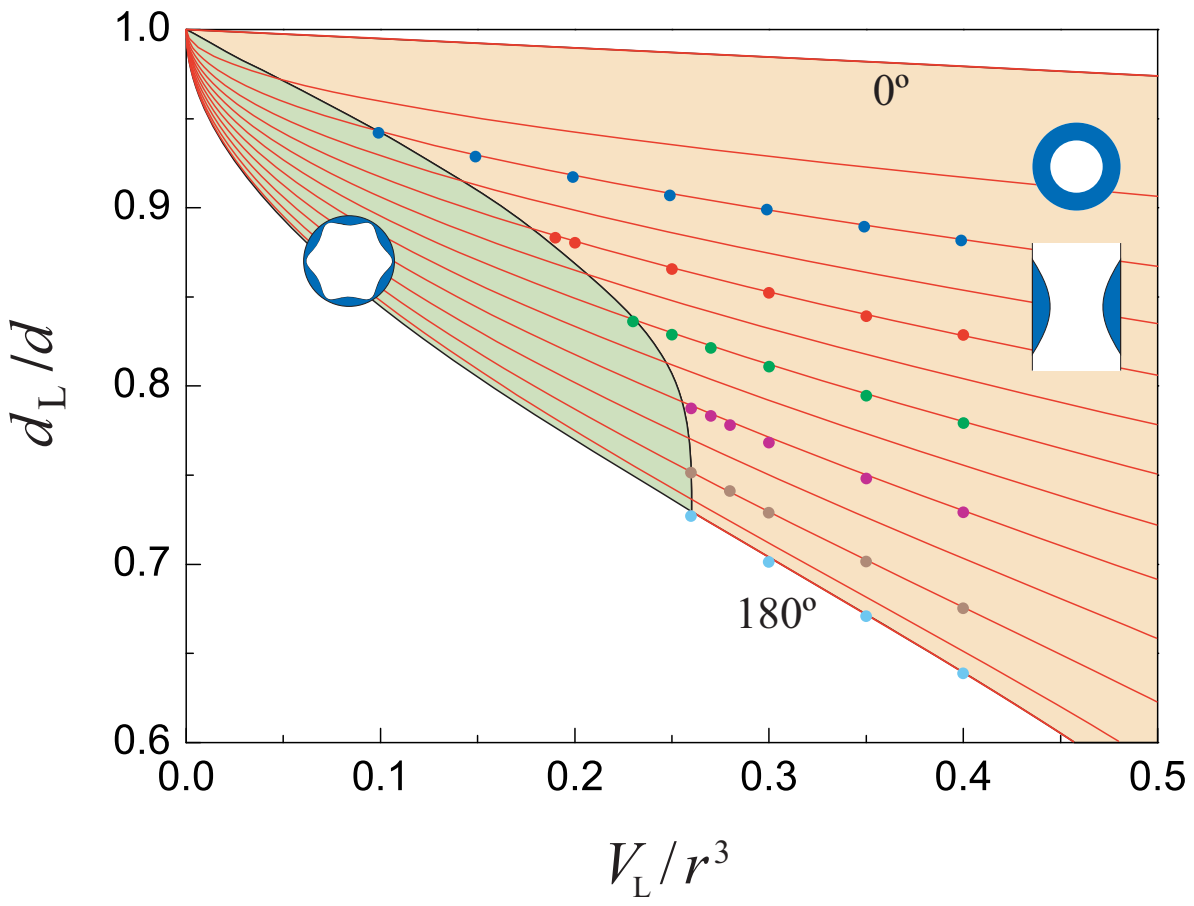

**Figure 4.** Enlarged region of the ($d_L$,$V_L$) parameter space, containing the boundary between axisymmetric and non-axisymmetric liquid morphologies. The light brown regions indicate axisymmetric liquid rings, the green regions non-axisymmetric morphologies resulting from the decay of a slender liquid ring into droplets. The corresponding morphologies are included as schematics. The red curves indicate the theoretical results, the dots the data from the *Surface Evolver* simulations. The black line is the boundary between axisymmetric and non-axisymmetric configurations.

A re-organization of the data from Figure 3 and Figure 4 enables us to represent the landscape of the possible liquid morphologies in a different way, as shown in Figure 5. In Figure 5(a), the light brown region with the upper (blue curve) and lower (red curve) boundaries shows the domain of stable axisymmetric wetting states in the ($d_L$, $\theta$) parameter space. The yellow region indicates the domain where axisymmetric plugs are the stable equilibrium configuration, the green region the domain where axisymmetry is broken due to the decay of a slender liquid ring into droplets. Figure 5(b) shows the corresponding regions in the ($V_L$, $\theta$) parameter space. The red dots represent $d_{Lc}$ or $V_{Lc}$ calculated via Eq. (27) or Eq. (28), and the blue squares are data points obtained using *Surface Evolver*, corresponding to the black curve in Figure 4.

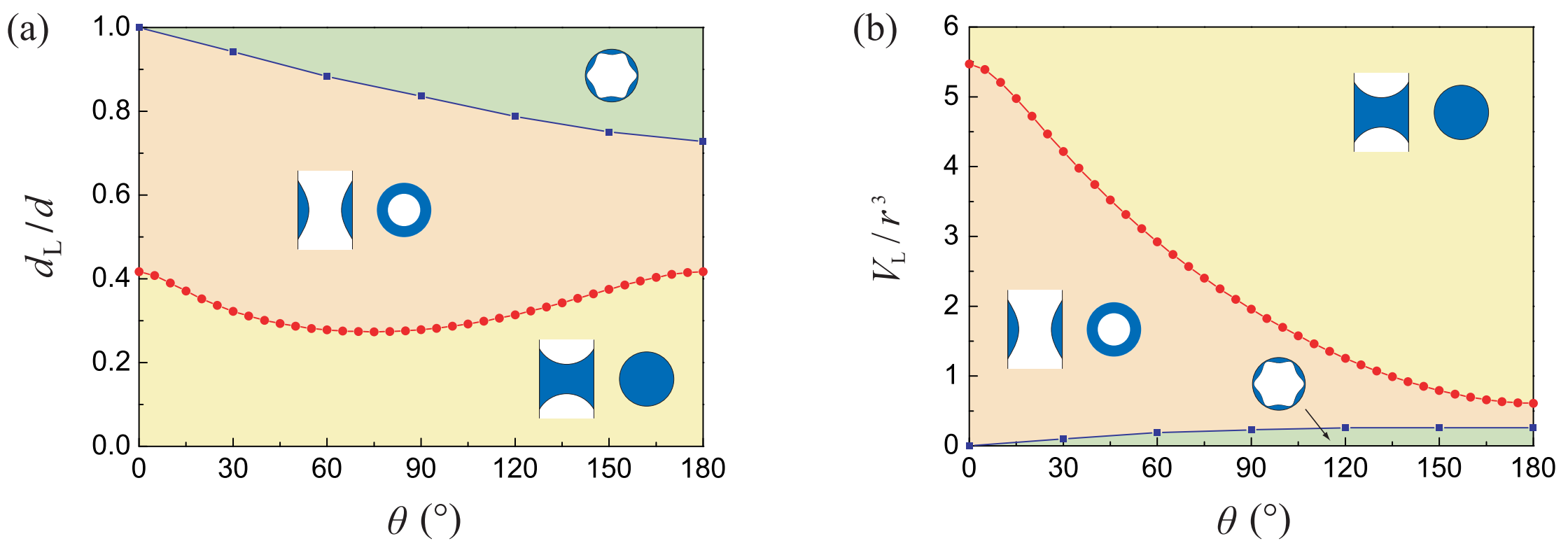

**Figure 5.** Possible liquid morphologies in a capillary tube. (a) Represented in ($d_L$, $\theta$) space. (b) Represented in ($V_L$, $\theta$) space. The schematic drawings indicate the types of morphology found in the different regions.

## CONCLUSIONS AND OUTLOOK

In summary, we systematically investigated the wetting morphologies of a liquid volume in a capillary tube, based on analytical calculations and numerical simulations. Based on an analytical solution of the axisymmetric Young-Laplace equation, we have determined the surface profile of a liquid ring for arbitrary contact angles. Moreover, based on a stability analysis we have determined the criteria for the transition from a liquid ring to a liquid plug. To the authors' best knowledge, this is the first time to report the complete set of critical parameters for this transition, i.e., the critical Laplace pressure, throat diameter, volume, solid-liquid contact length, etc. The analytical calculations were complemented by numerical simulations, especially for cases where the axial symmetry is broken. When comparing the theoretical and the numerical results for the axisymmetric case, a very good agreement is found. The non-axisymmetric equilibrium configurations found in the numerical simulations result from the decay a slender liquid ring via the Rayleigh-Plateau instability. The complete details of the non-axisymmetric decay require extensive future numerical and theoretical analysis, especially since effects due to contact-angle hysteresis come into play.

Finally, we would like to emphasize that our theory is expected to apply to very recent work about capillary instabilities of fluid threads in confinement (27) where oil-water systems were considered. In this work, no analytical solution for the interface shape was reported. Dimensionless critical parameters were obtained using various numerical methods, either applying to the static case or to the quasi-static case at very low capillary numbers. These values are very close to our analytical results for $\theta = 0°$. Therefore, we believe that our theory may provide a general theoretical framework also for such two-phase flows at small Reynolds and capillary numbers.

**AUTHER INFORMATION**

**Corresponding Author**

*E-mail: cunjinglv@tsinghua.edu.cn

**ORCID**

Cunjing Lv: 0000-0001-8016-6462

Steffen Hardt: 0000-0001-7476-1070

**Notes**

The authors declare no competing financial interest.

**ACKNOWLEDGEMENTS**

C.L. would like to gratefully acknowledge the support from the Alexander von Humboldt Foundation of Germany, the National Natural Science Foundation of China (Grant No. 11872227, 11632009), and Tsinghua University (Grant No. 53330100319), Beijing, China.

## Appendix A: Details related to the analytical solution of the Young-Laplace equation

Here, we provide the details for the calculations appearing in the main text. In particular, the derivatives of the interfacial energy and liquid ring volume (i.e. $\partial U/\partial r_{\mathrm{L}}$ and $\partial V_{\mathrm{L}}/\partial r_{\mathrm{L}}$) with respect to $r_{\mathrm{L}}$ are computed, as well as $\partial V_{\mathrm{L}}/\partial \Delta P$.

For the sake of simplicity, in what follows we employ the following dimensionless variables

$$\bar{r}_{\mathrm{L}} = \frac{r_{\mathrm{L}}}{r}, \quad 2\bar{H} = 2Hr, \quad \Delta\bar{P} = \Delta P \cdot \left(\frac{r}{\sigma}\right), \quad \bar{U} = \frac{U}{4\pi r^2 \sigma}, \quad \bar{V}_{\mathrm{L}} = \frac{V_{\mathrm{L}}}{2\pi r^3}. \tag{A1}$$

On the basis of Eq. (A1), we rewrite Eqs. (7), (18) and (19) as

$$2\bar{H} = \Delta\bar{P} = 2\left(\frac{\cos\theta - \bar{r}_{\mathrm{L}}}{1-\bar{r}_{\mathrm{L}}^2}\right), \tag{A2}$$

$$\begin{aligned}\bar{U} &= \left(\frac{1}{a}\right)\left(\frac{1}{a}+\bar{r}_{\mathrm{L}}-\cos\theta\right)E(\varphi,k)\Big|_{\varphi_1}^{\varphi_2=\frac{\pi}{2}} - \bar{r}_{\mathrm{L}}\cos\theta\, F(\varphi,k)\Big|_{\varphi_1}^{\varphi_2=\frac{\pi}{2}} \\ &= \left(\frac{1}{a}\right)\left(\frac{1}{a}+\bar{r}_{\mathrm{L}}-\cos\theta\right)\left[E(k)-E(\varphi,k)\big|_{\varphi_1}\right] - \bar{r}_{\mathrm{L}}\cos\theta\left[K(k)-F(\varphi,k)\big|_{\varphi_1}\right],\end{aligned} \tag{A3}$$

$$\begin{aligned}\bar{V}_{\mathrm{L}} &= \left[\frac{1}{a}E(\varphi,k)+\bar{r}_{\mathrm{L}}F(\varphi,k)\right]\Bigg|_{\varphi_1}^{\varphi_2=\frac{\pi}{2}} \\ &\quad -\frac{1}{3a}\left\{\left[\frac{2}{a^2}+\frac{3}{a}\bar{r}_{\mathrm{L}}+2\bar{r}_{\mathrm{L}}^2\right]E(\varphi,k)\Big|_{\varphi_1}^{\varphi_2=\frac{\pi}{2}} - \bar{r}_{\mathrm{L}}^2F(\varphi,k)\Big|_{\varphi_1}^{\varphi_2=\frac{\pi}{2}} - \sqrt{1-a^2}\sqrt{1-\bar{r}_{\mathrm{L}}^2}\right\} \\ &= \frac{1}{a}\left[E(k)-E(\varphi,k)\big|_{\varphi_1}\right]+\bar{r}_{\mathrm{L}}\left[K(k)-F(\varphi,k)\big|_{\varphi_1}\right] \\ &\quad -\frac{1}{3a}\left\{\left[\frac{2}{a^2}+\frac{3}{a}\bar{r}_{\mathrm{L}}+2\bar{r}_{\mathrm{L}}^2\right]\left[E(k)-E(\varphi,k)\big|_{\varphi_1}\right]-\bar{r}_{\mathrm{L}}^2\left[K(k)-F(\varphi,k)\big|_{\varphi_1}\right] \right. \\ &\quad \left. -\sqrt{1-a^2}\sqrt{1-\bar{r}_{\mathrm{L}}^2}\right\}\end{aligned}. \tag{A4}$$

The variables $a$ and $k$ can be rewritten as $a=\left(\cos\theta-\overline{r}_{\mathrm{L}}\right)/\left(1-\overline{r}_{\mathrm{L}}\cos\theta\right)$ and $k=\sqrt{1-\left(a\overline{r}_{\mathrm{L}}\right)^2}$. Since for a specific value of $\theta$, $a=\left(\overline{r}_{\mathrm{L}}\right)$, $k=\left(\overline{r}_{\mathrm{L}}\right)$ and $\varphi=\left(\overline{r}_{\mathrm{L}}\right)$ are solely functions of $\overline{r}_{\mathrm{L}}$ (note that from the main text we know $\varphi$ = arcsin[(1 – $a^2$)/$k^2$]$^{1/2}$), we obtain

$$\frac{\partial a}{\partial \overline{r}_{\mathrm{L}}}=-\frac{\sin^2\theta}{\left(1-\overline{r}_{\mathrm{L}}\cos\theta\right)^2}, \tag{A5}$$

$$\frac{\partial k}{\partial \overline{r}_{\mathrm{L}}}=-\frac{a\overline{r}_{\mathrm{L}}}{k}\cdot\frac{a-\overline{r}_{\mathrm{L}}}{1-\overline{r}_{\mathrm{L}}\cos\theta}, \tag{A6}$$

$$\frac{\partial \varphi}{\partial \overline{r}_{\mathrm{L}}}=\frac{1}{k^4\sin\varphi\cos\varphi}\cdot\frac{a}{\left(1-\overline{r}_{\mathrm{L}}\cos\theta\right)^2}\cdot\left(1-\overline{r}_{\mathrm{L}}^2\right)^2\sin^2\theta. \tag{A7}$$

Moreover, for the complete elliptic integrals $K(k)$ and $E(k)$ we have

$$\frac{\partial K(k)}{\partial \overline{r}_{\mathrm{L}}}=\frac{\partial K(k)}{\partial k}\cdot\frac{\partial k}{\partial \overline{r}_{\mathrm{L}}}=\left[\frac{E(k)}{k\left(1-k^2\right)}-\frac{K(k)}{k}\right]\cdot\frac{\partial k}{\partial \overline{r}_{\mathrm{L}}}, \tag{A8}$$

$$\frac{\partial E(k)}{\partial \overline{r}_{\mathrm{L}}}=\frac{\partial E(k)}{\partial k}\cdot\frac{\partial k}{\partial \overline{r}_{\mathrm{L}}}=\left[\frac{E(k)-K(k)}{k}\right]\cdot\frac{\partial k}{\partial \overline{r}_{\mathrm{L}}}. \tag{A9}$$

Considering that both $k(r_{\mathrm{L}})$ and $\varphi(r_{\mathrm{L}})$ in the incomplete elliptic integrals $F(\varphi, k)$ and $E(\varphi, k)$ are functions of $r_{\mathrm{L}}$, we obtain the derivatives $\mathrm{d}F\left(\varphi,k\right)/\mathrm{d}\overline{r}_{\mathrm{L}}$ and $\mathrm{d}E\left(\varphi,k\right)/\mathrm{d}\overline{r}_{\mathrm{L}}$ based on the following steps

$$\frac{\partial F(\varphi,k)}{\partial k}=-\frac{k\sin\varphi\sqrt{1-\sin^2\varphi}}{\left(1-k^2\right)\sqrt{1-k^2\sin^2\varphi}}-\frac{F(\varphi,k)}{k}+\frac{E(\varphi,k)}{k\left(1-k^2\right)}, \tag{A10}$$

$$\frac{\partial F(\varphi,k)}{\partial \varphi}=\frac{1}{\sqrt{1-k^2\sin^2\varphi}}, \tag{A11}$$

$$\frac{\mathrm{d}F(\varphi,k)}{\mathrm{d}\overline{r}_{\mathrm{L}}}=\frac{\partial F(\varphi,k)}{\partial k}\cdot\frac{\partial k}{\partial \overline{r}_{\mathrm{L}}}+\frac{\partial F(\varphi,k)}{\partial \varphi}\cdot\frac{\partial \varphi}{\partial \overline{r}_{\mathrm{L}}}, \tag{A12}$$

and

$$\frac{\partial E(\varphi,k)}{\partial k}=\frac{E(\varphi,k)-F(\varphi,k)}{k}, \tag{A13}$$

$$\frac{\partial E(\varphi,k)}{\partial \varphi}=\sqrt{1-k^2\sin^2\varphi}, \tag{A14}$$

$$\frac{\mathrm{d}E(\varphi,k)}{\mathrm{d}\bar{r}_\mathrm{L}}=\frac{\partial E(\varphi,k)}{\partial k}\cdot\frac{\partial k}{\partial \bar{r}_\mathrm{L}}+\frac{\partial E(\varphi,k)}{\partial \varphi}\cdot\frac{\partial \varphi}{\partial \bar{r}_\mathrm{L}}. \tag{A15}$$

Substituting Eqs. (A6), (A7), (A10) and (A11) into Eq. (A12), we obtain the expressions for $\mathrm{d}F(\varphi,k)/\mathrm{d}\bar{r}_\mathrm{L}$, and similarly for $\mathrm{d}E(\varphi,k)/\mathrm{d}\bar{r}_\mathrm{L}$.

After that, a combination of Eq. (A3) and Eqs. (A5) - (A15) leads to

$$\frac{\mathrm{d}\bar{U}}{\mathrm{d}\bar{r}_\mathrm{L}}=\alpha_1E(k)+\alpha_2K(k)+\alpha_3E(\varphi,k)+\alpha_4F(\varphi,k)+\alpha_5, \tag{A16}$$

in which the coefficients $\alpha_i$ ($i$ = 1, 2, … 5) are

$$\alpha_1=\frac{1}{a}+\left(\frac{2}{a^3}+\frac{\bar{r}_\mathrm{L}}{a^2}-\frac{\cos\theta}{a^2}\right)\cdot\frac{\sin^2\theta}{\left(1-\bar{r}_\mathrm{L}\cos\theta\right)^2}+\left(\frac{a-\bar{r}_\mathrm{L}}{1-a\bar{r}_\mathrm{L}}\right), \tag{A17}$$

$$\alpha_2=-\cos\theta+\frac{\bar{r}_\mathrm{L}}{1-a\bar{r}_\mathrm{L}}\cdot\left(\frac{a-\bar{r}_\mathrm{L}}{1-\bar{r}_\mathrm{L}\cos\theta}\right)\cdot\left(\frac{1-a\cos\theta}{a}\right), \tag{A18}$$

$$\alpha_3=-\alpha_1, \tag{A19}$$

$$\alpha_4=-\alpha_2, \tag{A20}$$

$$\alpha_5=\frac{1}{ak^2}\cdot\frac{\sqrt{1-\bar{r}_\mathrm{L}^2}}{\left(1-\bar{r}_\mathrm{L}\cos\theta\right)^3\sqrt{1-a^2}}\cdot\left(2\cos^2\theta-2\bar{r}_\mathrm{L}\cos\theta+\bar{r}_\mathrm{L}^2-1\right). \tag{A21}$$

Moreover, we obtain $\mathrm{d}\bar{V}_\mathrm{L}/\mathrm{d}\bar{r}_\mathrm{L}$ by a combination of Eq. (A4) and Eqs. (A5) - (A15):

$$\frac{\mathrm{d}\bar{V}_{\mathrm{L}}}{\mathrm{d}\bar{r}_{\mathrm{L}}} = \beta_1 E\left(k\right) + \beta_2 K\left(k\right) + \beta_3 E\left(\varphi, k\right) + \beta_4 F\left(\varphi, k\right) + \beta_5 , \tag{A22}$$

in which the coefficients $\beta_i$ ($i = 1, 2, \dots 5$) are

$$\begin{aligned} \beta_1 = & -\left(\frac{1}{a^2} + \frac{4\bar{r}_{\mathrm{L}}}{3a}\right) + \frac{\sin^2\theta}{\left(\cos\theta - \bar{r}\right)^2} \cdot \left(1 - \frac{2}{3}\bar{r}_{\mathrm{L}}^2 - \frac{2}{a^2} - \frac{2\bar{r}}{a}\right) \\ & - \frac{1}{k^2} \cdot \left(\frac{a - \bar{r}_{\mathrm{L}}}{\cos\theta - \bar{r}_{\mathrm{L}}}\right) \cdot \left(1 + a\bar{r} - \frac{\bar{r}}{3a} - \bar{r}_{\mathrm{L}}^2 - \frac{2}{3}a\bar{r}_{\mathrm{L}}^3\right) \end{aligned}, \tag{A23}$$

$$\beta_2 = 1 + \frac{2\bar{r}}{3a} + \frac{\bar{r}_{\mathrm{L}}^2 \sin^2\theta}{3\left(\cos\theta - \bar{r}\right)^2} + \frac{a\bar{r}}{k^2} \cdot \left(\frac{a - \bar{r}_{\mathrm{L}}}{\cos\theta - \bar{r}_{\mathrm{L}}}\right) \cdot \left(1 - \frac{2}{3a^2} - \frac{\bar{r}}{a} - \frac{\bar{r}_{\mathrm{L}}^2}{3} + a\bar{r}\right), \tag{A24}$$

$$\beta_3 = -\beta_1, \tag{A25}$$

$$\beta_4 = -\beta_2, \tag{A26}$$

$$\begin{aligned} \beta_5 = & -\frac{\bar{r}_{\mathrm{L}}}{3a} \cdot \frac{\sqrt{1 - a^2}}{\sqrt{1 - \bar{r}_{\mathrm{L}}^2}} \\ & + \frac{\sin^2\theta}{\left(1 - \bar{r}\cos\theta\right)^3} \cdot \frac{\sqrt{1 - \bar{r}_{\mathrm{L}}^2}}{3a^2 k^2 \sqrt{1 - a^2}} \cdot \left[\left(1 - a^2\bar{r}_{\mathrm{L}}^2\right) \cdot \left(1 - \bar{r}\cos\theta\right) + \left(1 - \bar{r}_{\mathrm{L}}^2\right) \cdot \left(-3a^2 + 2 + 3a\bar{r} + 2a^2\bar{r}_{\mathrm{L}}^2\right)\right] \\ & - \left(\bar{r} + \frac{\bar{r}_{\mathrm{L}}^2}{3a}\right) \cdot \frac{1}{ak^2} \cdot \left(\frac{1}{1 - \bar{r}\cos\theta}\right) \cdot \left[\frac{\left(a - \bar{r}_{\mathrm{L}}\right) \cdot \sqrt{1 - a^2} \cdot \sqrt{1 - \bar{r}_{\mathrm{L}}^2}}{\bar{r}} + \frac{\left(1 - \bar{r}_{\mathrm{L}}^2\right)^2 \sin^2\theta}{\left(1 - \bar{r}\cos\theta\right)^2 \sqrt{1 - a^2} \cdot \sqrt{1 - \bar{r}_{\mathrm{L}}^2}}\right] \end{aligned} \tag{A27}$$

On the basis of Eq. (A2) and Eq. (A22) we obtain

$$\frac{\mathrm{d}\bar{V}_{\mathrm{L}}}{\mathrm{d}\Delta\bar{P}} = \frac{\mathrm{d}\bar{V}_{\mathrm{L}} / \mathrm{d}\bar{r}_{\mathrm{L}}}{\mathrm{d}\Delta\bar{P} / \bar{r}_{\mathrm{L}}}, \tag{A28}$$

in which

$$\frac{\mathrm{d}\Delta\bar{P}}{\mathrm{d}\bar{r}_{\mathrm{L}}}=\frac{\mathrm{d}\left(2\bar{H}\right)}{\mathrm{d}\bar{r}_{\mathrm{L}}}=\frac{-2\left(1+\bar{r}_{\mathrm{L}}^{2}-2\bar{r}_{\mathrm{L}}\cos\theta\right)}{\left(1-\bar{r}_{\mathrm{L}}^{2}\right)^{2}}. \tag{A29}$$

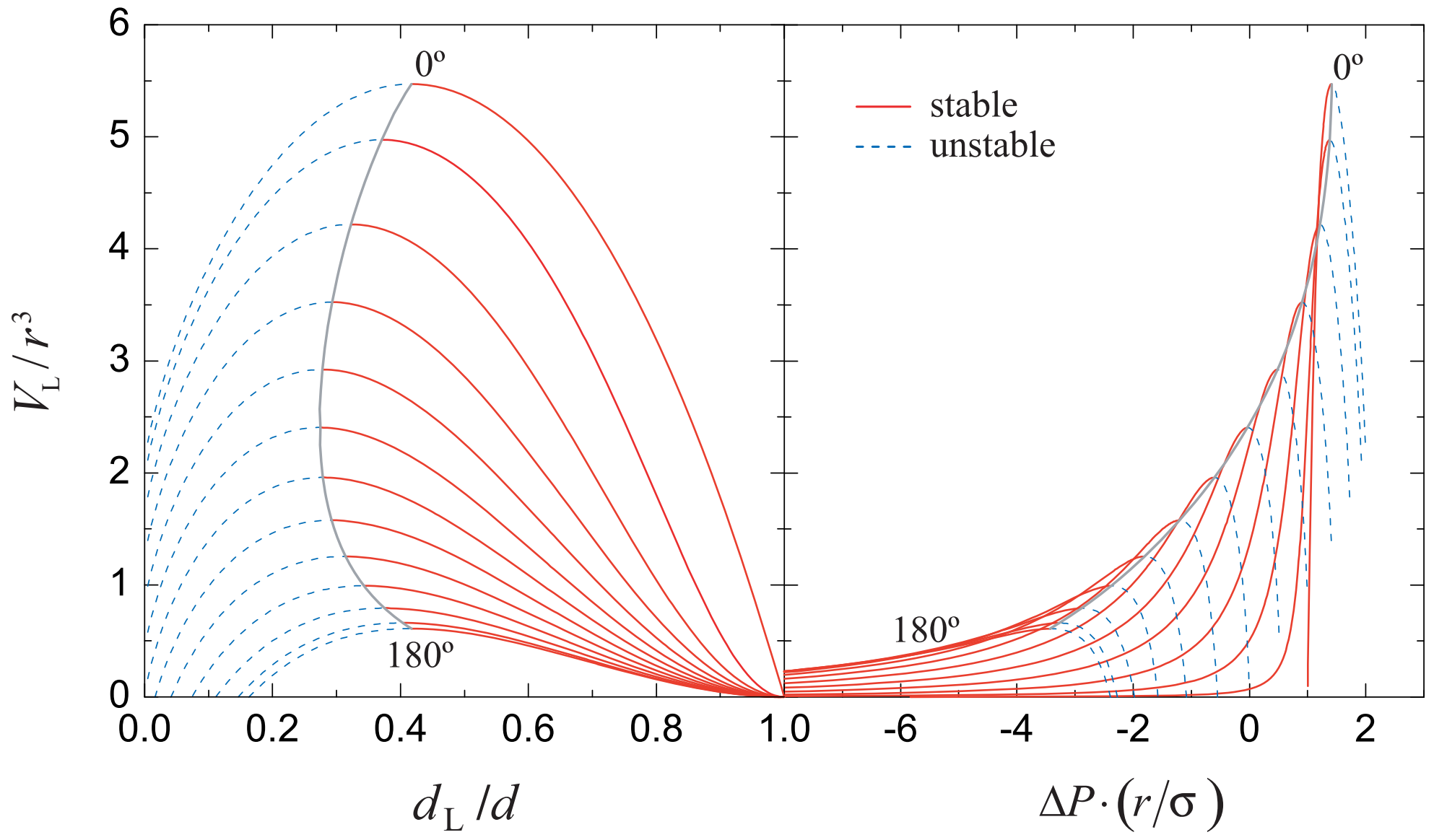

**Figure A1.** Dependence of $V_{\mathrm{L}}$ on $d_{\mathrm{L}}$ and $\Delta P$ expressed in dimensionless form. The contact angle ranges from 0° to 180°, with increments of 15°. The solid red and dashed blue curves represent the stable and unstable regimes, respectively, with the solid grey curves representing the boundaries between the stable and unstable regime.

In Figure A1, we display the dependence of $V_{\mathrm{L}}$ on $d_{\mathrm{L}}$ and $\Delta P$ in dimensionless form. The right panel of Figure A1 shows that for each $\theta$, $V_{\mathrm{L}}$ first increases with $\Delta P$ but then decreases. At the same time, $\Delta P$ monotonically decreases with increasing $d_{\mathrm{L}}$. This confirms the existence of the critical point ($\Delta P_{\mathrm{c}}$, $V_{\mathrm{Lc}}$) accounting for the instability (i.e. the point where $\mathrm{d}V_{\mathrm{L}}/\mathrm{d}\Delta P = 0$). For a specific $\theta$, the values of $V_{\mathrm{Lc}}$ obtained in the left and right panels are exactly identical.

In Figure A2, the blue dots, red hollow circles, and green crosses represent the result of $\mathrm{d}\bar{U}/\mathrm{d}\bar{r}_{\mathrm{L}}=0$, $\mathrm{d}\bar{V}_{\mathrm{L}}/\mathrm{d}\bar{r}_{\mathrm{L}}=0$ and $\mathrm{d}\bar{V}_{\mathrm{L}}/\mathrm{d}\Delta\bar{P}=0$ on the basis of Eqs. (A16), (A22) and (A28), respectively. They are perfectly consistent with each other. However, when $\theta\in$ [0°, 90°], $\mathrm{d}\bar{U}/\mathrm{d}\bar{r}_{\mathrm{L}}=0$ has another solution branch, which is plotted using blue dots and a blue curve as a guide to the eye. Analyzing the relevance of this solution branch, however, is beyond the scope of this paper and is left to future work. The red dots plotted in Figure 5 of the main text is the same data as that of the main solution branch shown in Figure A1.

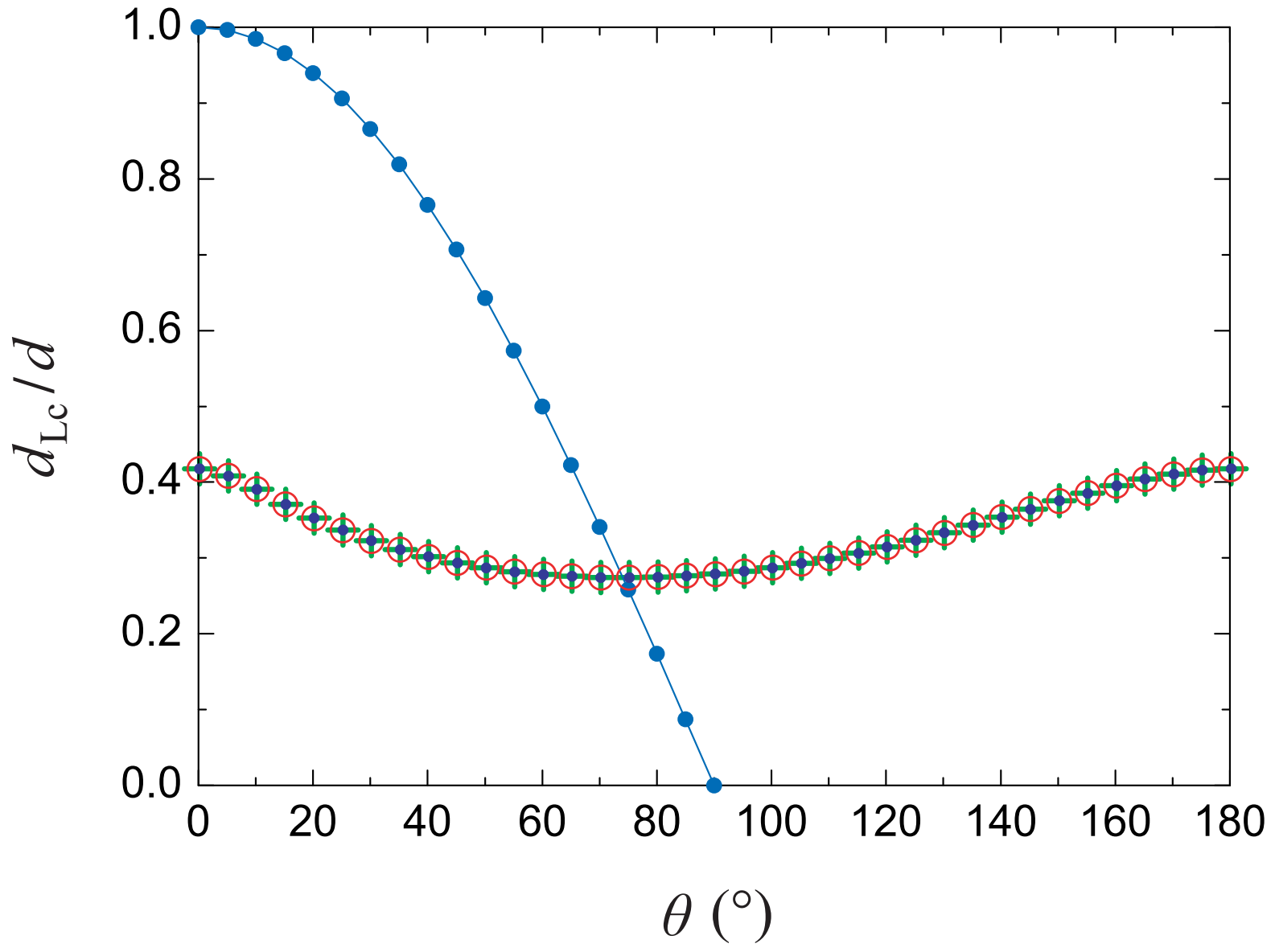

**Figure A2.** Dependence of the dimensionless critical liquid throat diameter $d_{\mathrm{Lc}}$ on $\theta$, resulting from $\mathrm{d}V_{\mathrm{L}}/\mathrm{d}r_{\mathrm{L}} = 0$ (red hollow circles), $\mathrm{d}U/\mathrm{d}r_{\mathrm{L}} = 0$ (blue solid circles) and $\mathrm{d}V_{\mathrm{L}}/\mathrm{d}\Delta P = 0$ (green crosses). Also shown is the second solution branch of $\mathrm{d}U/\mathrm{d}r_{\mathrm{L}} = 0$ (blue dots, connected by lines).

When $\theta = 0°$ we obtain $a = 1$, $k = \sqrt{1-\bar{r}_{\mathrm{L}}^2}$, $\varphi_1 = 0$, $E(\varphi_1, k) = 0$, $F(\varphi_1, k) = 0$, and $\alpha_i$, $\beta_i$ ($i$ = 1, 2, … 5) become $\alpha_1 = 2$, $\alpha_2 = -1$, $\alpha_3 = -2$, $\alpha_4 = 1$, $\alpha_5 = 0$, $\beta_1 = -2\left(1+\bar{r}_{\mathrm{L}}\right)$, $\beta_2 = 1+\bar{r}_{\mathrm{L}}$, $\beta_3 = 2\left(1+\bar{r}_{\mathrm{L}}\right)$, $\beta_4 = -\left(1+\bar{r}_{\mathrm{L}}\right)$, $\beta_5 = 0$. With that Eqs. (A16), (A22) and (A28) lead to

$$\frac{\mathrm{d}\bar{U}}{\mathrm{d}\bar{r}_{\mathrm{L}}} = 2E\left(k\right) - K\left(k\right), \tag{A30}$$

$$\frac{\mathrm{d}\bar{V}_{\mathrm{L}}}{\mathrm{d}\bar{r}_{\mathrm{L}}} = -\left(1+\bar{r}_{\mathrm{L}}\right)\left[2E\left(k\right) - K\left(k\right)\right], \tag{A31}$$

$$\frac{\mathrm{d}\bar{V}_{\mathrm{L}}}{\mathrm{d}\Delta\bar{P}} = \frac{1}{2}\left(1+\bar{r}_{\mathrm{L}}\right)^3\left[2E\left(k\right) - K\left(k\right)\right]. \tag{A32}$$

Finally, $\mathrm{d}\bar{U}/\mathrm{d}\bar{r}_{\mathrm{L}} = 0$, $\mathrm{d}\bar{V}_{\mathrm{L}}/\mathrm{d}\bar{r}_{\mathrm{L}} = 0$ and $\mathrm{d}\bar{V}_{\mathrm{L}}/\mathrm{d}\Delta\bar{P} = 0$ result in

$$\left[2E(k) - K(k)\right] = 0. \tag{A33}$$

These results lead to the following critical parameters for the onset of the instability: $k = 0.909$, $2H_c r = 1.411$, $d_{Lc}/d = 0.417$, $U_c/(r^2\sigma) = -6.081$, $V_c/r^3 = 5.471$ and $h_{Lc}/r = 4.257$, etc.

**Appendix B: Details of the Surface Evolver simulations**

To verify our theoretical results on the equilibrium and stability of the axisymmetric liquid ring and the criteria for the potential morphology transition into non-axisymmetric configurations, we resort to FEM simulations. The public domain software package *Surface Evolver* (SE) (33) was employed to compute the static configurations of the liquid surface.

The basic concept of SE is to minimize the energy and find the equilibrium shape of a liquid volume surface with given surface tension, subjected to external forces (e.g. gravity, centrifugal force, magnetic force) and constraints (e.g. volume conservation, contact angle, pinning of the contact line). SE has been applied to studying various wetting phenomena (36-40), with excellent agreement to experimental results. However, SE cannot not deal with dynamic problems.

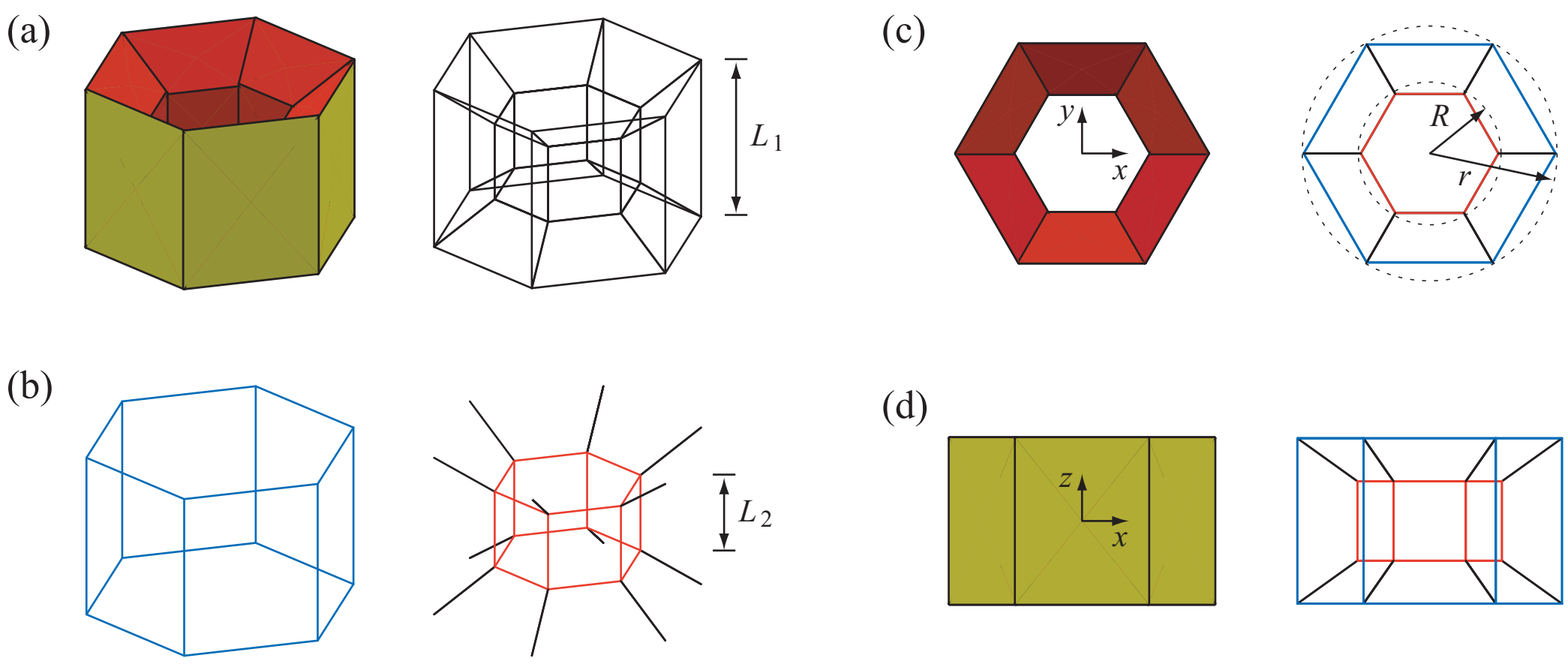

**Figure A3.** Initial configuration of the liquid ring in the capillary tube. Shown is the computational grid. For better visualization, different colors are used to denote the facets and the edges. The red and yellow areas represent the liquid-vapor and solid-liquid interfaces, respectively. The blue and red edges represent the outer and the inner regular hexagonal surfaces, respectively. The outer and the inner hexagonal surface have a same geometric center. (a) Liquid ring from an oblique side view; (b) Outer (blue) and inner (red) surface of the liquid ring. (c) Liquid ring from a top view. (d) The same from a side view. Components of the edges of the liquid ring. The relevant geometric parameters $L_1$, $L_2$, $R$ and $r$ are indicated.

SE build up geometric objects from planar facets. This means that also the cylindrical tube needs to be approximated by planar facets. This is done based on successively finer approximations. The simulations require an initial configuration. To speed up the calculation and to promote convergence, ideally we choose an initial configuration that is not too far away from the final one. As shown in Figure A3 from different views, we define an initial configuration by employing two regular hexagonal surfaces with the same geometric center. The enveloped volume is the liquid ring, and in the simulations this volume is fixed. To focus on the liquid surface, the wall of the tube is not shown. With $r$ being the inner radius of the tube, we need to define suitable initial values of $L_1$, $L_2$ and $R$ to determine the initial configuration. $R$ is the radius of the circumcircle of the inner hexagon, $L_1$ and $L_2$ is the length of the outer and inner hexagonal surface, respectively. Since for given values of $\theta$ and $V_L$, we can obtain $h_L$ and $r_L$ using the theory given in the main text, we assign $L_1 = h_L$ and $R = r_L$. Generally, we can choose an arbitrary value for $L_2$, here we assign $L_2 = 0.6\ L_1$. The liquid volume of the initial configuration (Figure A3(a)) does not necessarily need to be equal to $V_L$, because SE will automatically correct it during the iteration process.

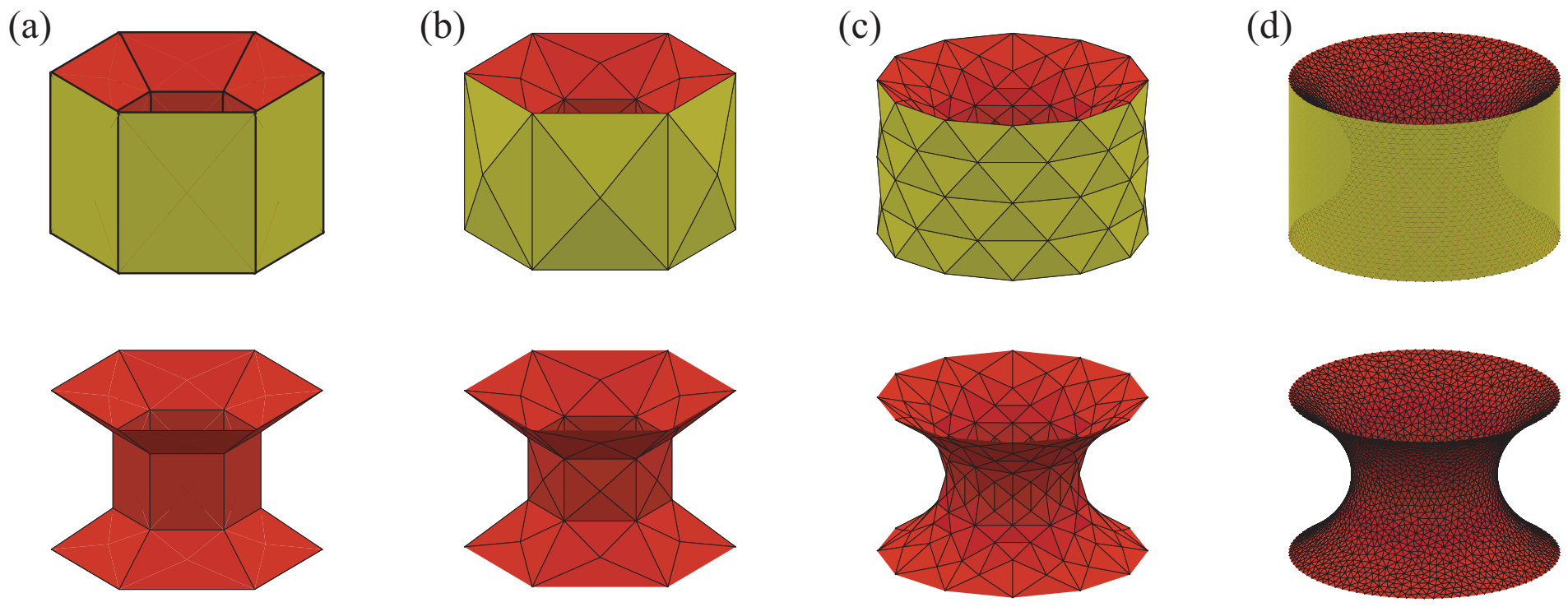

**Figure A4.** Configurations of the liquid ring at different stages of the iteration process. (a) Initial configuration (b), (c) Intermediate configurations. (d) Equilibrium state after a sufficient number of iterations, corresponding to $\theta$ = 60°, $d_L/d$ = 0.546, $h_L/r$ = 1.251, $2Hr$ = – 0.131, $V/r^3$ = 2.0 and $U/(r\sigma)$ = 3.356. For better visualization, the corresponding liquid-vapor interfaces (red) with the computational grid are shown separately.

In the simulations, we define a constraint to the solid-liquid interface, i.e., a cross section through this interface needs to be described by $x^2 + y^2 = r^2$. This constraint is fulfilled with increasing accuracy during the iterations when the circle is approximated by polygons. The solid-liquid-vapor three-phase contact line can move along the axial direction.

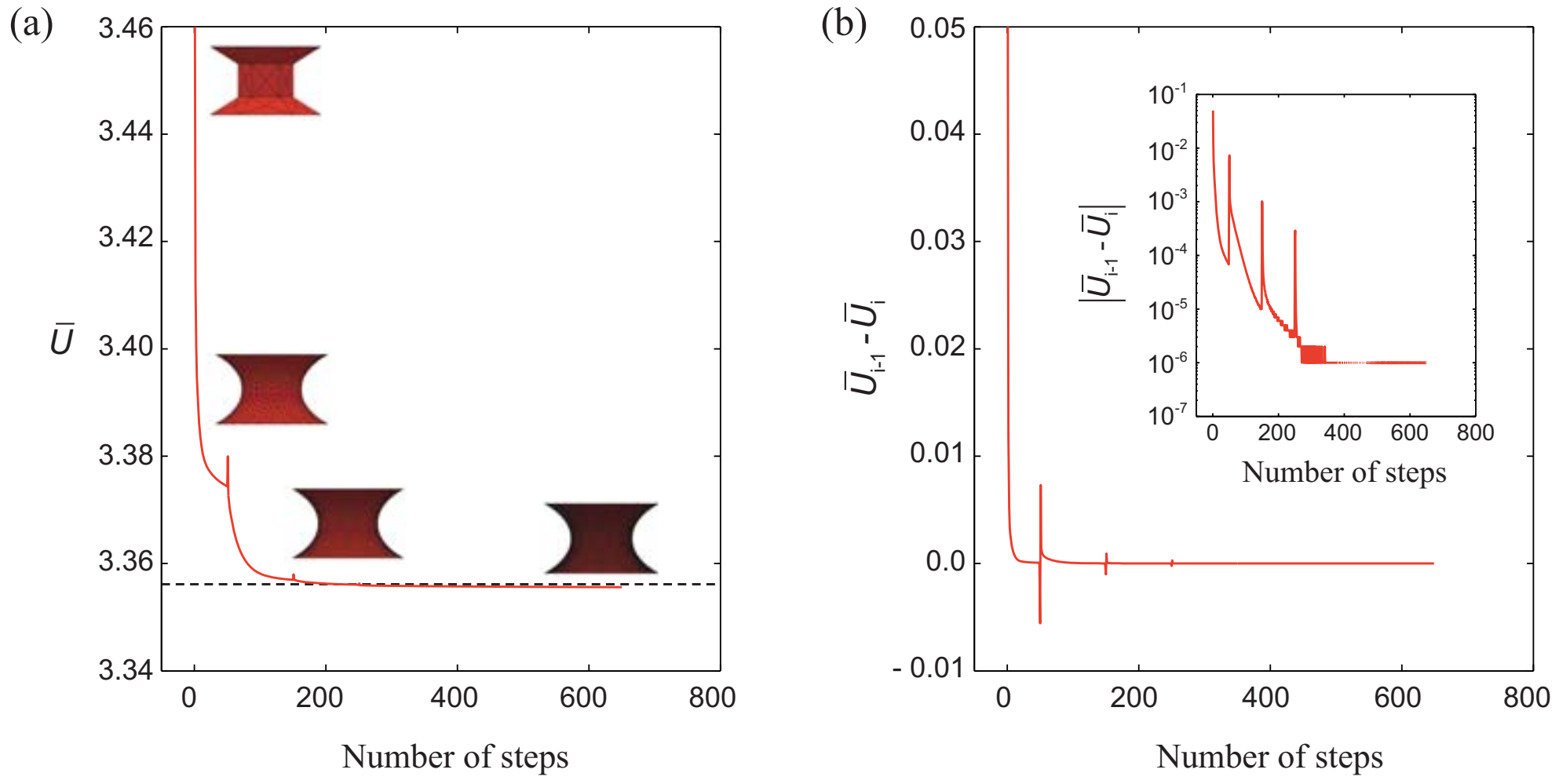

**Figure A5.** (a) Iteration process of SE. The red line represents the evolution of the dimensionless

surface energy. The black dashed line is the theoretical result. The insets are the configurations of the liquid ring at different stages of the iteration process. (b) Evolution of the dimensionless quantities $U_{i-1} - U_i$ and $|U_{i-1} - U_i|$.

During a simulation, SE computes the surface energy of the system $U = \sigma(A_{LV} - A_{SL}\cos\theta)$ after each iteration. After several iterations, the liquid surface becomes smooth. The surface energy of the system decreases as the iterations progress and finally approaches an asymptotic value. We denote this value $U_0$. To find $U_0$, we define a tolerance interval $\delta_i = |U_{i-1} - U_i|/(r^2\sigma)$, denoting $U_i$ the surface energy at the $i$th iteration (see Figure A5). When $\delta_i < 10^{-5}$, we suppose we are close enough to the asymptotic state, i.e. $U_{i-1} \approx U_i \approx U_0$. Typically, in our simulations convergence is reached after several hundred iterations with $10^4 - 10^5$ triangular facets.